\documentclass[conference]{IEEEtran}
\IEEEoverridecommandlockouts

\usepackage{color}
\usepackage{soul}
\usepackage{ulem}
\usepackage{cite}
\usepackage{threeparttable}
\usepackage{amsmath,amssymb,amsfonts}
\usepackage{algorithmic}
\usepackage{graphicx}
\usepackage{textcomp}
\usepackage{xcolor}
\usepackage{tabularx}
\usepackage{url}
\usepackage{booktabs}
\usepackage{svg}
\usepackage[colorlinks, linkcolor=blue]{hyperref}

\def\BibTeX{{\rm B\kern-.05em{\sc i\kern-.025em b}\kern-.08em
    T\kern-.1667em\lower.7ex\hbox{E}\kern-.125emX}}
\begin{document}

\title{Longitudinal Wrist PPG Analysis for Reliable Hypertension Risk Screening Using Deep Learning
}

\author{\IEEEauthorblockN{Hui Lin\IEEEauthorrefmark{1}\IEEEauthorrefmark{3},
Jiyang Li\IEEEauthorrefmark{1}, 
Ramy Hussein\IEEEauthorrefmark{1},
Xin Sui\IEEEauthorrefmark{1}, 
Xiaoyu Li\IEEEauthorrefmark{2}, 
Guangpu Zhu\IEEEauthorrefmark{2},\\
Aggelos K. Katsaggelos\IEEEauthorrefmark{3},
Zijing Zeng\IEEEauthorrefmark{2}, and
Yelei Li\IEEEauthorrefmark{2}
\IEEEauthorblockA{\IEEEauthorrefmark{1}OPPO Health Lab, Palo Alto, CA 94303, USA}
\IEEEauthorblockA{\IEEEauthorrefmark{2}OPPO Health Lab, Shenzhen 518037 China}
\IEEEauthorblockA{\IEEEauthorrefmark{3}Department of Electrical and Computer Engineering, Northwestern University, Evanston, IL 60208 USA}
\thanks{
Corresponding author: Ramy Hussein (email: ramy.hussein@oppo.com).}}
}




\maketitle

\begin{abstract}

Hypertension is a leading risk factor for cardiovascular diseases. Traditional blood pressure monitoring methods are cumbersome and inadequate for continuous tracking, prompting the development of PPG-based cuffless blood pressure monitoring wearables. This study leverages deep learning models, including ResNet and Transformer, to analyze wrist PPG data collected with a smartwatch for efficient hypertension risk screening, eliminating the need for handcrafted PPG features. Using the Home Blood Pressure Monitoring (HBPM) longitudinal dataset of 448 subjects and five-fold cross-validation, our model was trained on over 68k spot-check instances from 358 subjects and tested on real-world continuous recordings of 90 subjects. The compact ResNet model with 0.124M parameters performed significantly better than traditional machine learning methods, demonstrating its effectiveness in distinguishing between healthy and abnormal cases in real-world scenarios.

\end{abstract}


\section{Introduction}

Hypertension is a leading risk factor for cardiovascular, cerebrovascular, and renal diseases, affecting an estimated 1.28 billion adults globally. Due to its often asymptomatic nature, approximately 46\% of individuals with hypertension are unaware of their condition \cite{who_hypertension}. Early detection prevents severe health complications and promotes effective health management and lifestyle interventions. However, traditional blood pressure measurement relies on bulky cuff-based devices, which can be inconvenient and uncomfortable for frequent monitoring. These single-point measurements also fail to capture blood pressure fluctuations throughout the day. Recent advances in sensor technology and the widespread use of smartwatches have enabled photoplethysmography (PPG) signals for continuous, cuff-less blood pressure monitoring. This innovative approach provides a more comfortable and continuous means of tracking blood pressure, enhancing hypertension risk screening and overall health management.



Recent studies have shown promising results in hypertension detection using PPG data. These approaches typically rely on extracting morphological features from PPG signals and their derivatives, which are then used as input for classification models such as support vector machines and random forests \cite{he2022new}. While these methods perform well with high-quality PPG signals, their accuracy degrades significantly when the signals are affected by motion artifacts, breathing, or other types of interference. Such noise distorts the PPG waveform, making it difficult to extract features that reliably differentiate between various blood pressure categories. To overcome this limitation, deep learning models have been proposed to process raw PPG data directly, which automatically learn robust features, reduce the reliance on handcrafted features that are sensitive to noise, and potentially lead to more accurate and reliable hypertension detection \cite{10360188,qin2023cuff,truong2022non,paviglianiti2022comparison}.



Several studies have explored deep-learning methods for hypertension detection and blood pressure estimation. For instance, Schlesinger~\textit{et al.}~\cite{schlesinger2020blood} proposed convolutional and Siamese neural network models for PPG-based blood pressure measurement, achieving an average regression performance with a mean absolute difference of 5.95 mmHg for systolic BP and 3.41 mmHg for diastolic BP. Similarly, Suhas~\textit{et al.}~\cite{suhas2024end} introduced a Transformer-based model with a weighted contrastive loss, achieving an average mean absolute error of 1.08 mmHg for systolic BP (SBP) and 0.68 mmHg for diastolic BP (DBP), respectively. Vraka1~\textit{et al.}~\cite{vraka2023harnessing} use the Inception-ResNet-v2 network for hypertension detection, achieving a sensitivity of 72.8\% and specificity of 76.2\%. While these methods demonstrate strong results, they are typically validated on curated, high-quality PPG datasets collected from fingertips while people are sitting still, such as MIMIC~\cite{johnson2016mimic} and PulseDB~\cite{wang2023pulsedb}, which do not reflect the noise and variability present in real-world data. These models often struggle when faced with noisy PPG signals distorted by motion artifacts and other interferences. Moreover, their performance is impacted by the covariate shift problem, where variations in PPG data across different populations lead to significant degradation in accuracy. As a result, the reliability and generalizability of these approaches in real-world settings remain uncertain.

To address these challenges, our contributions are outlined as follows:
\begin{itemize} 
\item Broad Participant Base: We collected PPG data and corresponding blood pressure readings from 448 subjects, representing a diverse demographic range in terms of age, weight, height, and gender. This diversity enhances the generalizability of our findings across various applications and populations.
\item Longitudinal Analysis: We assess hypertension risk using PPG signals collected over a long period rather than single-point measurements. By requiring multiple effective days and analyzing trends over time, we enhance the reliability of hypertension detection.
\item Advanced Modeling: After preprocessing the PPG signals, we deployed ResNet architectures with varying model sizes. Our approach significantly outperforms other methods with a compact model of just 0.124M parameters, demonstrating the robustness and efficiency of deep learning for wrist-based hypertension detection.
\item Real-World Performance Evaluation: Trained on benchmark blood pressure and PPG data collected intermittently when an Omron device was available, our model was tested on PPG data gathered continuously in the background during regular daily activities, which includes dynamic and noisy nature of physiological signals in varied real-world conditions. The evaluation validates our model’s robustness and practicality for everyday usage.
\end{itemize}


\section{Methods}

\begin{figure}[htb]
    \centering
    \includegraphics[width=.45\textwidth]
    {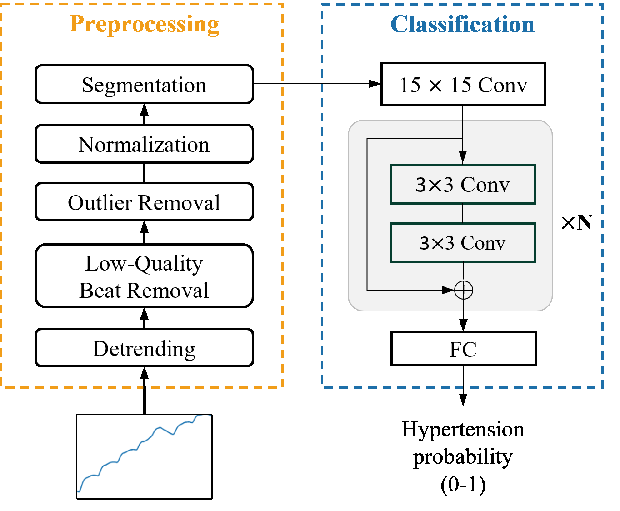}
    \caption{
    Overview of the end-to-end deep learning framework for hypertension detection. The framework consists of two key components: (1) PPG preprocessing, which includes detrending, low-quality beat removal, outlier removal, normalization, and segmentation, and (2) a ResNet-based feature learning and classification model that processes the preprocessed PPG data to predict hypertension risk, where Conv denotes convolutional layers, $N$ represents the number of residual blocks, and FC refers to fully connected layers.}
    
    \label{fig:overall}
\end{figure}

The end-to-end deep learning framework for hypertension detection is illustrated in Fig.~\ref{fig:overall}. It consists of two main components: PPG preprocessing and a ResNet-based feature learning and classification model. The preprocessing stage includes steps such as PPG detrending, low-quality beat removal, outlier detection, normalization, and signal segmentation, ensuring the PPG data is properly prepared for the classification phase. In the next stage, a ResNet architecture is used to automatically learn discriminative features from the preprocessed PPG data and predict the probability of hypertension.


\begin{figure}[htb]
    \centering
     \includegraphics[width=0.5\textwidth]
    {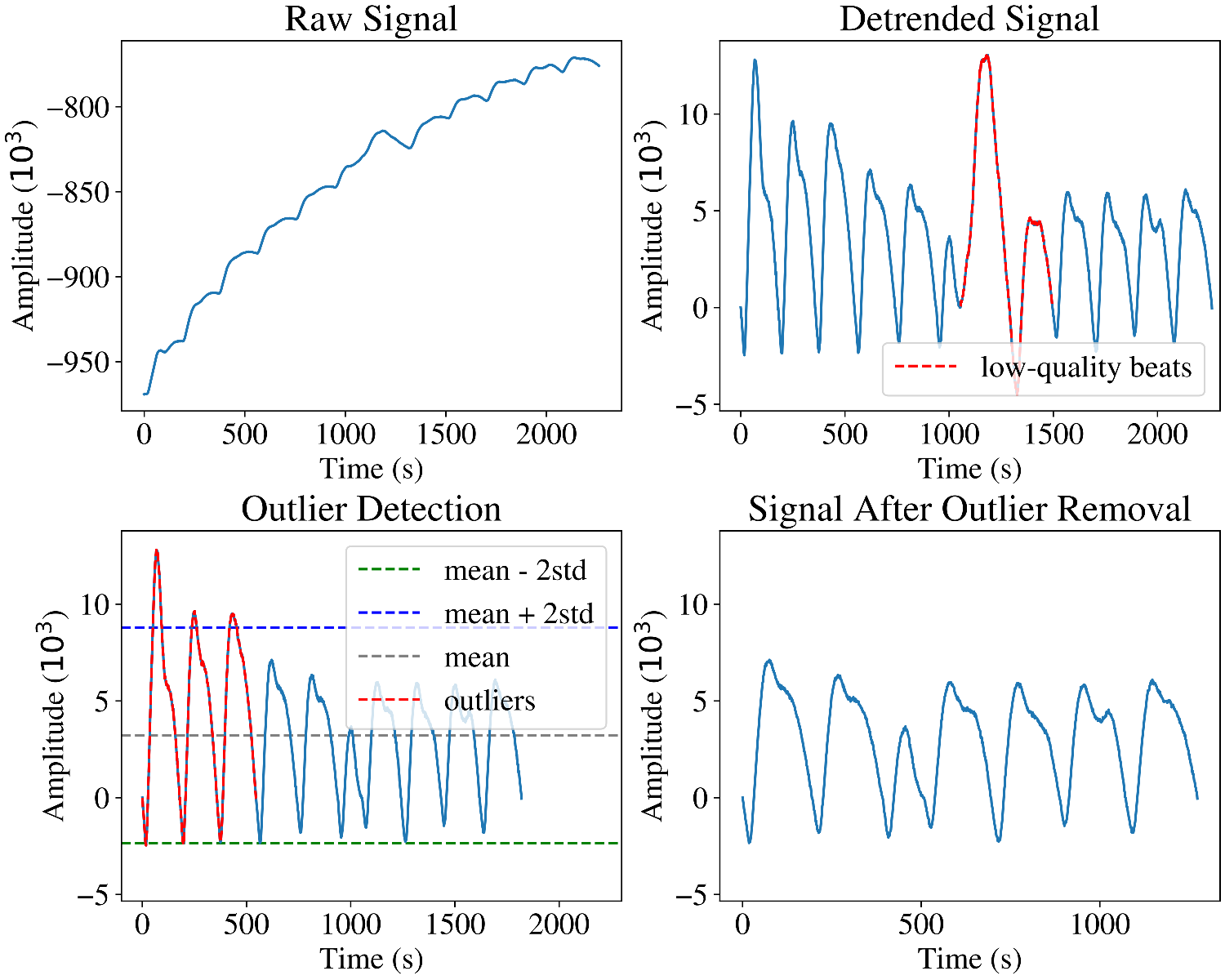}
    \caption{Overview of the preprocessing pipeline for PPG signals. The raw PPG signal (top left) shows a baseline drift, which is removed in the detrended version (top right) where low-quality beats are identified in red. The bottom left panel illustrates outlier detection based on z-scores, with outliers marked in red outside the blue and green boundaries. The bottom right panel presents the signal after outlier removal, which is prepared for further processing.
    }
    \label{fig:preprocessing}
\end{figure}

\subsection{PPG Preprocessing}
The signal preprocessing pipeline is essential for improving the quality and reliability of PPG data. Ensuring it is suitable for analysis and modeling. As shown in Fig.~\ref{fig:overall}, the pipeline comprises five key steps, each addressing specific challenges in raw PPG signals:\\
1) Detrending: This step corrects baseline drift caused by factors such as respiration or sensor movement. By removing this drift, the focus shifts to critical beat-to-beat changes, improving the accuracy of subsequent analyses. The upper left plot of Fig.~\ref{fig:preprocessing} illustrates an example of baseline drift.\\
2) Low-Quality Beat Removal: This step filters out beats affected by noise or artifacts, ensuring that only high-quality data is retained for analysis. Using sliding window template matching, standard heartbeat templates are generated and compared against the signal, with beats scoring below a threshold classified as low quality. The upper right plot of Fig.~\ref{fig:preprocessing} highlights these identified artifacts.\\
3) Outlier Removal: Beats with z-scores outside the range of $[-2, 2]$ are discarded to mitigate the impact of anomalies, improving data consistency. The bottom left plot of Fig.~\ref{fig:preprocessing} marks outliers in red, with the blue and green dashed lines indicating the normal range. The bottom right plot shows the resulting clean PPG data.\\
4) Normalization: The PPG signals are scaled using min-max normalization to a range of [0, 1]. This step standardizes the data, facilitating more effective model training.\\
5) Segmentation: During training, a 4-second segment is randomly cropped from each signal as input. For testing, signals are divided into multiple 4-second segments, and the average of the predicted probabilities from all segments is used as the final prediction for the entire signal.


\subsection{Hypertension Detection}
\label{model}

After PPG preprocessing, we leverage a ResNet neural network architecture \cite{he2016deep} for feature learning and hypertension classification. The model begins with a convolutional layer using a 15x15 kernel, followed by $N$ iterations of 3x3 convolutional blocks forming the residual units, as shown in Fig.~\ref{fig:overall}. This is followed by two fully connected layers, which produce the final output—an estimated probability of hypertension ranging from 0 to 1. The model uses binary cross-entropy as the loss function, defined as:

\begin{equation}
    \mathcal{L}_{\text{CE}} = G \text{log}P + (1-G) \text{log}(1-P)
\end{equation}
where \(G\) is the ground truth, and \(P\) is the predicted probability. Note that $G$ is a binary value ($G\in \{0, 1\}$), while $P$ falls in the range of [0, 1]. 

We explored different values of $N$ ranging from 1 to 10, with the model achieving the highest ROC-AUC score when $N$ was set to 1. Thus, we take $N = 1$ (ResNet-4) and compare it with other models.


Additionally, downsampling of the majority class is applied to balance the datasets between healthy and abnormal cases, preventing biased representation and predictions.

\section{Experiments}
\label{Experiments}

\subsection{HBPM Dataset}

The Home Blood Pressure Monitoring (HBPM) dataset comprised 448 subjects, with demographic details presented in Table~\ref{table1}. Participants for the study were recruited from the specified medical centers. The inclusion criteria targeted males and females aged 20-60 with a BMI under 30. Exclusion criteria ruled out individuals who were on hypertension medication, diagnosed with pathological arrhythmias, peripheral artery-related diseases within the last 6 months, or using specific cardiovascular treatment devices. 


This study was conducted in Guangdong, Hebei, and Sichuan in collaboration with West China Hospital of Sichuan University~(IRB no. 54/2023) and Fuwai Hospital Chinese Academy of Medical Sciences, Shenzhen~(IRB no. SP2022112(01)A), adhered to the Declaration of Helsinki principles and only used deidentified physiological data. The study received approval from the institutional review boards (IRBs) at each center. All participants provided informed consent by local IRB requirements at enrollment. And researchers couldn't access any personal identifying information of the participants.


The study utilizes two distinct data types: labeled spot-check data collected at rest, and unlabeled background signals recorded without moving restrictions. For labeled spot-check data, we collected 1-minute PPG data and corresponding blood pressure readings three times daily, each preceded by a minute of rest, using the OPPO Watch~3~Pro (paired green sensors; sample frequency 250~Hz) and a digital cuff-based Omron BP monitor. The unlabeled background data was collected by the same wearable device continuously every 30 minutes throughout the day. For hypertension risk screening, a subject is labeled as ``positive'' when the average systolic blood pressure~(SBP) exceeds 120~mmHg or the average diastolic blood pressure~(DBP) exceeds 80~mmHg~\cite{AHA2024}. Our objective is to evaluate the model’s performance on background signals to reveal its performance of continuous health monitoring in real-world conditions.



\begin{table}[!t]
    \caption{Demographic description of subjects in the HBPM dataset.}
    \renewcommand\arraystretch{1.3}
    \centering
    \scalebox{1}{
    \small
    \begin{tabularx}{1\linewidth}{@{}X c@{}}
        \toprule[1pt]
        Item & HBPM Dataset \\
        \midrule      
        Subjects (M/F) & 448 (249/199)\\
        Signals (S/B) & 748493 (136907/611586) \\
        Abnormal Subjects & 189 \\
        Age (years) & 37.92 ± 8.59 (20-60) \\
        SBP (mmHg) & 111.07 ± 13.36 (79.76-151.60) \\
        DBP (mmHg) & 75.59 ± 9.61 (53.76-108.60) \\
        \bottomrule
    \end{tabularx}}
    \begin{tablenotes}
        \footnotesize
        \item M/F: Male/Female; S/B: Spot-check/Background.
        \item SBP: Systolic Blood Pressure; DBP: Diastolic Blood Pressure.
    \end{tablenotes}
    \label{table1}
\end{table}

       


\subsection{Model Validation}
\label{metrics}

We use five-fold cross-validation to ensure a robust evaluation of the model's performance across different subsets of the dataset.
Key performance metrics such as specificity, recall, sensitivity, ROC AUC (Area Under the Receiver Operating Characteristic Curve), and PR AUC (Precision-Recall Area Under the Curve) are essential for evaluating the effectiveness of the hypertension classification model. 
Additionally, the number of parameters of each model is compared to provide valuable insights into the trade-offs between model performance and computational demands.


The testing protocol: the validity of results for each subject depends on two key criteria. Firstly, only background samples are used for inference. Secondly, the positive ratio, calculated as the average of all probabilities, is used to determine the positive~(hypertension or prehypertension) probability for the subject.

\section{Results and Discussion}
\label{results} 


To assess the effectiveness of the proposed PPG preprocessing framework and ResNet, we compared its performance with both traditional machine learning models and other deep learning models. Table~\ref{table:performance_comparison}~shows the performance metrics of various models for hypertension risk screening. Specifically, ResNet-4 is compared against a computationally efficient LightGBM model based on a gradient-boosting machine framework, as well as a Transformer-based model \cite{vaswani2017attention}. For the LightGBM model, we use handcrafted PPG morphological features described as follows: Systolic and diastolic peaks are first identified in the PPG signals. Then other fiducial points are derived from the second derivative of the PPG signals. Features are calculated from relative differences in magnitude and time between these characteristic points. There are 15 selected features used in the LightGBM model to identify elevated blood pressure. The Transformer-based model includes a ResNet-based encoder with an 8-head, single-layer Transformer encoder for hypertension risk screening, processing 10 sequences of 1-second PPG segments.

\begin{table}[!t]
    \caption{Evaluation of Various Models on Background Signals}
    \renewcommand\arraystretch{1.5}
    \centering
    \small
    \setlength{\tabcolsep}{1mm}{
    \begin{tabularx}{1.0\linewidth}{@{}l c c c c c@{}}
        \toprule[1pt]
        Model & Precision & Sensitivity  & Specificity & ROC\_AUC & PR\_AUC \\
        \midrule
        LightGBM & 86.3\% & 46.6\% & 94.6\% & 84.1\% & 80.8\%\\
        ResNet-4 & 87.5\% & 44.4\% & 95.4\% & \textbf{86.5\%} & \textbf{83.0\%}\\
        Transformer & 89.6\% & 45.5\% & 96.1\% & \underline{86.0\%} & \underline{81.9\%}\\
        \bottomrule
    \end{tabularx}}
    \begin{tablenotes}
        \footnotesize
        \item \textbf{Bold}: Best results; \underline{Underline}: Second-best results.
    \end{tablenotes}
    \label{table:performance_comparison}
\end{table}

\begin{table}[!t]
    \caption{The Number of Parameters between Various Models}
    \renewcommand\arraystretch{1.5}
    \centering
    \small
    \begin{tabularx}{1.0\linewidth}{c c c c}
        \toprule[1pt]
        Model & LightGBM & ResNet-4  & Transformer \\
        \midrule
        Parameters~(MB) & \textbf{0.048} & \underline{0.124} &   4.674\\
        \bottomrule
    \end{tabularx}
    \begin{tablenotes}
        \footnotesize
        \item \textbf{Bold}: Best results; \underline{Underline}: Second-best results.
    \end{tablenotes}
    \label{table:size_comparison}
\end{table}

\begin{figure}
    \centering
    \includegraphics[width=.49\linewidth]
    {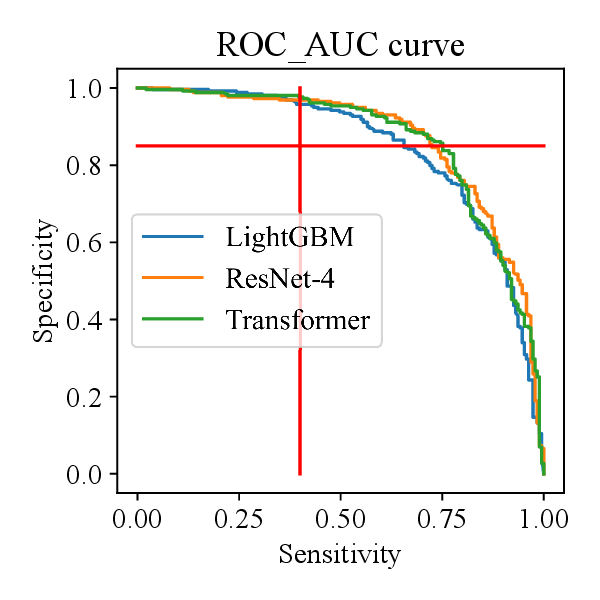}
    \includegraphics[width=.49\linewidth]{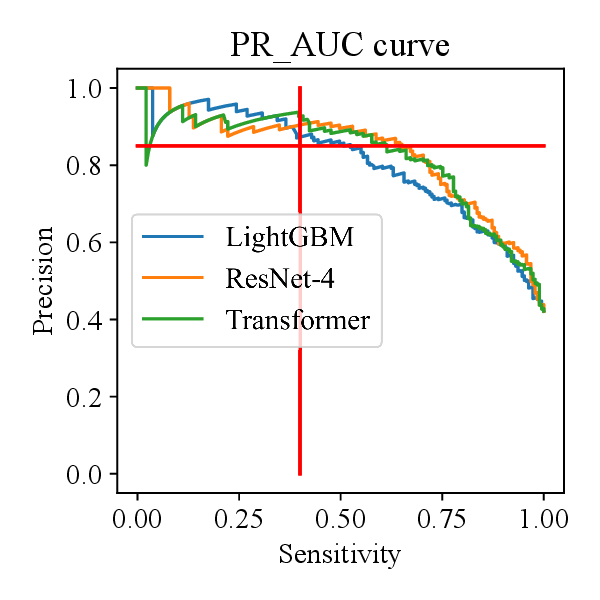}
    \caption{Comparison of ROC and PR curves for three models. The red line is a reference. The plot shows that the ResNet-4 outperforms in both metrics.
    }
    \label{fig:ROC}
\end{figure}

Notably, ResNet-4 achieves the highest ROC AUC score of 86.5\% and the highest PR AUC score of 83\%, indicating overall best performance across different levels of sensitivity. Fig. \ref{fig:ROC} presents the ROC and PR curves of our method compared to others. ResNet-4 also performs competitively in terms of precision and specificity, with values of 87.5\% and 95.4\%, respectively. The high specificity and precision enable the model to identify true positive abnormal cases with fewer false negatives and false positives, making it highly suitable for applications requiring precise hypertension risk screening and accurate classification of healthy subjects. 


Table \ref{table:size_comparison} compares the model sizes for three models. ResNet-4 offers the best performance with a size of 0.124 MB, the second smallest model, and far smaller than the transformer with a size of 4.674 MB. Therefore, ResNet-4 is ideal for its best overall performance and computational efficiency. The LightGBM model, at 0.048 MB, is the most compact for environments with strict size constraints. The Transformer model achieves a good performance, but its complex structure makes it more suitable for resource-rich scenarios.







\section{Conclusion}


This study has demonstrated the effectiveness of the ResNet-4 model in hypertension risk screening, achieving a strong balance between sensitivity and specificity. The model accurately identifies true positive cases of abnormal blood pressure while minimizing false positives, making it well-suited for real-world applications. By employing a simple yet efficient deep learning model on well-preprocessed wrist PPG data, this approach offers a reliable, non-invasive solution for continuous blood pressure monitoring. Looking ahead, we plan to expand our dataset to include a broader and more diverse participant pool, covering various ages, ethnicities, and health profiles. Additionally, we aim to validate the model in real-time deployment scenarios and optimize its energy efficiency for seamless integration into resource-constrained wearable devices, enhancing its practicality for continuous health monitoring.

\section*{Acknowledgment}
The authors would like to thank OMRON Healthcare for their generous provision of blood pressure monitoring devices, which greatly facilitated our data acquisition process. 

\bibliographystyle{IEEEtran}

\end{document}